# Dual bound states in the continuum in metamaterials

Longqing Cong, and Ranjan Singh[*]

*Division of Physics and Applied Physics, School of Physical and Mathematical Sciences, Nanyang Technological University, Singapore 637371, Singapore*
*Centre for Disruptive Photonic Technologies, The Photonics Institute, Nanyang Technological University, 50 Nanyang Avenue, Singapore 639798, Singapore*
[*]E-mail: ranjans@ntu.edu.sg

**Abstract:** Bound state in the continuum (BIC) is a mathematical concept with an infinite radiative quality factor ($Q$) that exists only in an ideal infinite array. It was first proposed in quantum mechanics, and extended to general wave phenomena such as acoustic, water, elastic, and electromagnetic waves. In photonics, it is essential to achieve high $Q$ resonances for enhanced light-mater interactions that could enable low-threshold lasers, ultrasensitive sensors, and optical tweezers. Here, we demonstrate *dual* bound states in the continuum in a subwavelength planar metamaterial cavity that reveal symmetry-protected features excited by orthogonal polarizations. The spectral features of dual BICs are experimentally verified in the terahertz domain by breaking the $C_2$ symmetry that invokes a leakage channel. The radiative $Q$ factors tend to infinity at a discrete symmetry-restoring point and obey an inverse square dependence on the structural asymmetry. Metamaterials allow field confinement in extremely small mode volumes, thereby improving the rate of spontaneous emission in the cavity with a much larger Purcell factor along with high $Q$ factor. The symmetry-protected BICs in metamaterials also possess the unique advantage of scalability at different wavelengths for potential applications in sensing, lasing, switching, and spectral filtering.

A partial or complete confinement of electromagnetic (EM) waves in a cavity is essential for versatile applications with strong light-matter interactions, such as lasing[1-3], nonlinearity[4-6], and sensors[7-9]. The most straightforward idea to trap light includes optical fibers or Fabry-Perot cavities in macroscopic space, and electrons bound to nucleus or trapped in a quantum well in microscopic space. However, these regular bound states are commonly outside the radiation continuum which have no access to radiation channels[10]. Recently, artificially designed resonator array opened up another avenue to access extreme confinement of photons into micro- or nanoscale region[11, 12] above the light line within the radiation continuum. In general, EM wave is usually interpreted in terms of frequency spectrum, and it propagates as a spectral continuum above the light line in different media. Different orders of resonances may be found in the propagating continuum at specific frequencies described by $\tilde{\omega}_n = \omega_{n0} + i\gamma_n$, where the real part indicates resonance frequencies and the imaginary part indicates the leakage rates of the *n*-order mode[13]. For a strongly confined mode, the leakage rate is significantly reduced, and goes to zero if photons are perfectly trapped as a so-called *bound state* in an ideal scenario[14]. However, such a *bound state in the continuum* (BIC) is not observable from the spectrum due to the nonradiative feature with a zero spectral linewidth. As an embedded eigenvalue of the photonic system[15], it becomes visible in the spectrum with a finite leakage rate as a quasi-BIC by introducing external perturbations. Such a BIC-inspired mechanism for light confinement allows a general strategy to access extremely high quality factor (*Q*, $Q = \omega_0/2\gamma$) resonance in optical cavities.

Dielectric photonic crystals (PhCs) are an excellent platform to explore the BIC which avoids the intrinsic Ohmic loss for extremely high *Q* generation by exciting displacement current[11, 16]. Nevertheless, the geometric size of unit cells in PhCs is usually at the scale of resonance wavelength, and resonant energy is thus loosely confined in the resonator cavity with a relatively large mode volume (*V*). In this condition, Purcell factor that is proportional to *Q/V* is largely reduced in the traditional PhCs despite of a large *Q*, which hinders the efficient spontaneous emission rates in the context of cavity quantum electrodynamics[17]. An efficient approach to densely confine photons is by using metamaterials whose building blocks are in the subwavelength scale, and the resonant mode volume could be further reduced to a deep subwavelength scale by carefully designing the resonator geometry and mode properties. Here, we demonstrate *dual symmetry-protected BICs* in a subwavelength metamaterial that are independently induced by orthogonal polarizations of incident light. The features of the subwavelength BICs are experimentally observed in the terahertz spectrum by slightly breaking the $C_2$ symmetry with large $Q$ factors. We also unveil the inverse square dependence of $Q$ factor on the structural asymmetry parameter which describes a generalized evolution trajectory of the symmetry-protected BIC system.

A typical subwavelength double-gap split ring resonator (DSRR) is designed as a unit cell as shown in Figure 1a with geometrical parameters indicated in the graph. An infinite planar array consisting of periodic DSRRs is essential to realize an ideal bound state without edge scattering. Without loss of generality, we analyze the eigenmodes by exploring the ratio between period and wavelength ($p/\lambda$) that clearly manifests the subwavelength nature of the BICs in DSRRs. Here, Ohmic loss is ignored by considering a perfect electric conductor (PEC) as constituent of DSRRs with length of the square resonator $l = 0.8p$, width $w = 0.08p$, and gap size $g = 0.04p$. The two split gaps are located at the center of top and bottom arms, and thus lead to a $C_2$ rotational symmetry and mirror symmetry. The periodic array with $C_2$ symmetric unit cells can only allow symmetry-compatible vectors of electric and magnetic fields in the normal direction (*z* direction, $k_{//} = (k_x, k_y) = (0,0)$ ) radiating to the far field. As observed in Figure 1b, the two eigenmodes of interest disappear at $\Gamma$ point due to symmetry incompatibility, and emerge at off $\Gamma$ point once the symmetry is broken. The error bars indicate the resonance linewidth (leakage rates), which vanishes at $\Gamma$ point for both eigenmodes. The two eigenmodes are excited by orthogonal polarizations along the symmetry axes, and possess the features of a *symmetry-protected bound state*[10]. According to the eigenmode analysis, the subwavelength DSRR array supports two ideal polarization-dependent, symmetry-protected BICs with infinite $Q$ factors at $\Gamma$ point that are unstable against perturbations of symmetry breaking. A slight perturbation induces leakage of an ideal bound state, and thus leads to a quasi-BIC that has observable spectral features with finite linewidth. For clarity, we define the two eigenmodes as quasi-BIC I and quasi-BIC II for TE and TM polarizations, respectively.

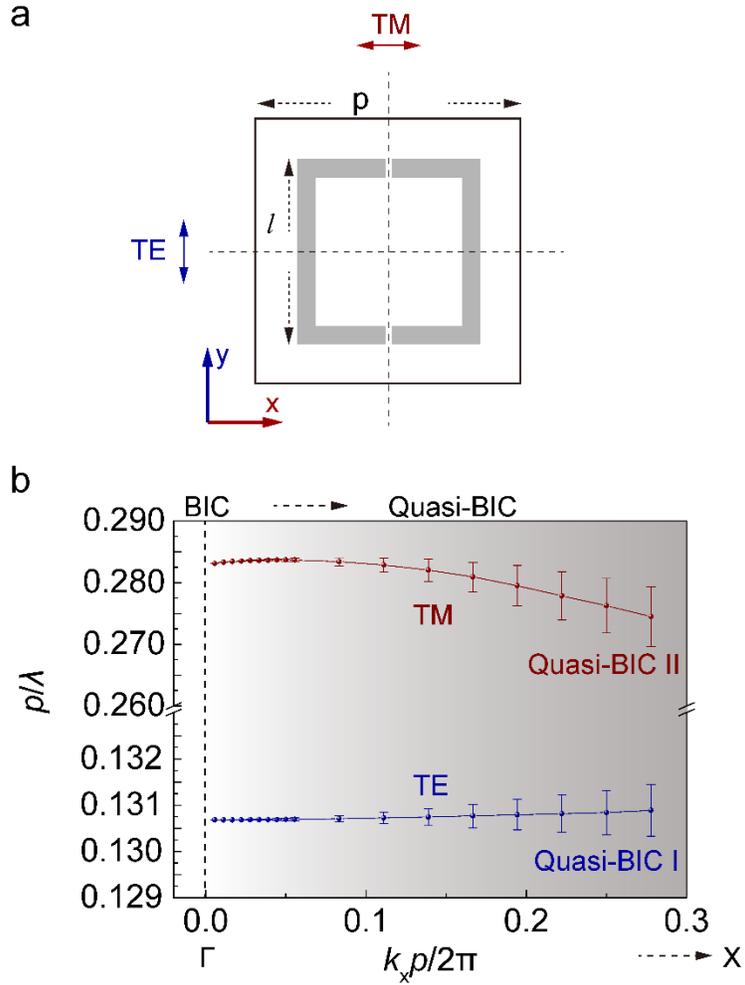

**Figure 1. Eigenmode analysis of a symmetric subwavelength resonator array. a.** Schematic diagram of a subwavelength resonator with inversion and $C_2$ symmetry as a building block of a metamaterial array. $C_2$ rotational symmetry leads to the anisotropic optical responses with TE- and TM-polarized excitations. The resonator is designed as a square ring with two split gaps, and the resonator length is indicated as $l$ residing on a square lattice with period of $p$. **b.** Eigenmodes with TE and TM excitations, with error bars indicating the linewidth of eigenmodes. Only the eigenmodes of interest are shown for clarity. All the resonance energy is completely confined as a bound state in the radiation continuum at $\varGamma$ point that are defined as the BICs.

For an intuitive interpretation of the BICs, numerical simulations based on finite element method are executed by sweeping the angle of incidence wavevector ($\theta$) relative to the infinitely extended resonator plane as shown in Figure 2a. Although the simulations in this work has been carried out in the terahertz regime, the scalability of metamaterials makes it possible to extend the discussion into other frequency bands. To stress the subwavelength feature of the BICs and be consistent with eigenmode analysis, we present the resonance frequencies in terms of $p/\lambda$ relative to the incident wavevector angle as shown in Figures 2b and 2c. We observe symmetric patterns relative to axis of $\theta = 0$ and the disappearance of resonance features at a discrete point of $\theta = 0$ ($\Gamma$ point) for the two quasi-BICs, which verify the eigenmode analysis of an ideal bound state. A genuine BIC is a strictly mathematical concept with an infinite radiative $Q$ factor that exists only in an ideal infinite array[18]. We extracted the total $Q$ factors instead of radiative $Q$ factors since Ohmic loss is excluded ($\gamma_{ohm} = 0$) in simulations with PEC ($Q_{tot} = \omega/2\gamma_{tot} = \omega/2(\gamma_{rad} + \gamma_{ohm}) = Q_{rad}$). Here, $Q_{tot}$ is the total quality factor comprising of radiative ($Q_{rad}$) and Ohmic ($Q_{Ohm}$) quality factors ($\frac{1}{Q_{tot}} = \frac{1}{Q_{Rad}} + \frac{1}{Q_{Ohm}}$). In order to obtain the leakage rate, we use Fano line shape equation $T_{Fano} = \left| a_1 + ja_2 + \frac{b}{\omega - \omega_0 + j\gamma_{tot}} \right|^2$ to fit the transmission intensity spectra, where $a_1$, $a_2$, and $b$ are real numbers, $\omega_0$ is the real part of central frequency of the quasi-BICs, and $\gamma_{tot}$ is the total leakage rate[19]. In practice, an ideal BIC is not observable since resonance linewidth vanishes without any dominant spectral feature, and a general strategy is to trace the diverging trajectory of $Q$ factor from a quasi-BIC scenario. As shown in Figures 2d and 2e, the $Q$ factors reveal a trend to diverge to infinity at $\theta = 0$ for both modes that manifests as the quasi-BIC features.

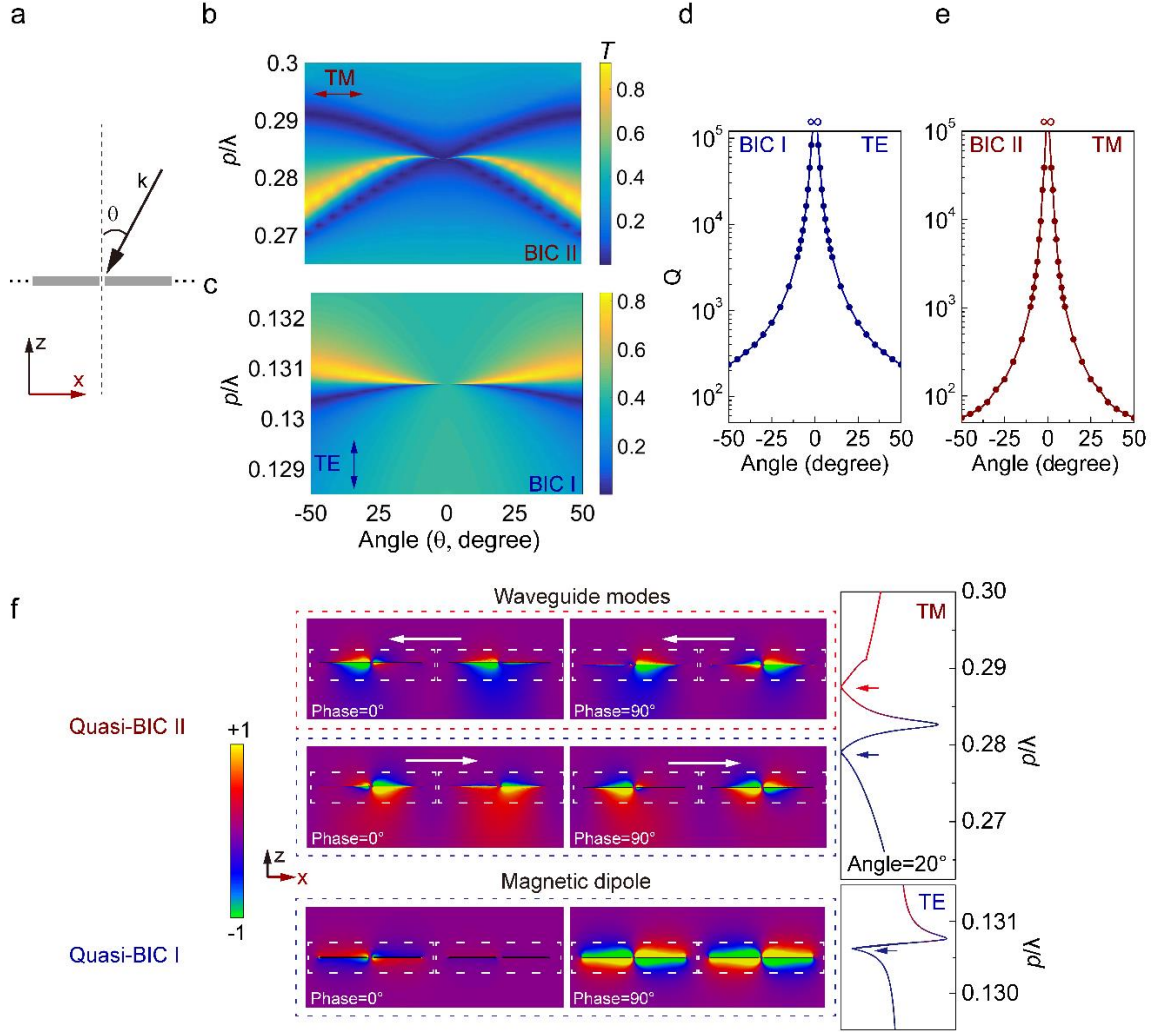

**Figure 2. Spectral analysis of quasi-BICs. a.** Pictorial illustration showing the angle-resolved spectral analysis by sweeping the incident wavevector angle $\theta$. **b, c.** Transmission spectra of BIC I and BIC II by sweeping the incident angle from -50 ° to 50 °. **d, e.** Extracted quality factors of the quasi-BICs. At discrete points above the light line, $Q$ tends to infinity indicating a bound state in the ideal subwavelength resonator array. **f.** Analysis of normalized $z$-polarized electric field distribution of Quasi-BIC I and Quasi-BIC II at an incident angle of 20 °. Waveguide modes propagating in the opposite directions at the transmission dips open an EIT-like quasi-BIC with TM polarization, and the neighboring antiphase oscillating electric dipoles are excited accounting for the emergence of the Fano-like quasi-BIC at TE polarization. The dashed rectangles indicate the unit cell area.

Although similar quasi-BIC features are observed in the polarization-dependent DSRR metamaterial at TE and TM polarizations, they basically originate from different mechanisms. The simulated surface electric field distributions clearly present the mode patterns related to the two quasi-BICs in Figure 2f. At TM polarization, the quasi-BIC II is formed due to the emergence of waveguide modes propagating in opposite directions at the two spectral dips (top panel of Figure 2f, more clearly visualized in the supplementary video). The two waveguide modes can couple to in-plane wavevectors ($k_{//}$), and open an leakage channel that forms an electromagnetic induced transparency (EIT) like[20] mode at $\theta \neq 0$. On the other hand, two antiphase oscillated electric dipoles on the two branches of a resonator couples to radiation due to slight phase mismatch for quasi-BIC II (see the bottom panel of Figure 2f at phase = 0°, and also the supplementary video). This phase mismatch breaks the symmetry of the mode pattern, and thus leads to a leakage manifesting itself as a Fano-like spectrum[21] at $\theta \neq 0$. These two modes are originally uncoupled to radiation due to incompatibility to the perfect $C_2$ symmetry at $\theta = 0$. The antiphase oscillated electric fields on the neighboring branches of an individual resonator form a closed current loop flowing on the DSRR that also enhances the magnetic dipole moment.

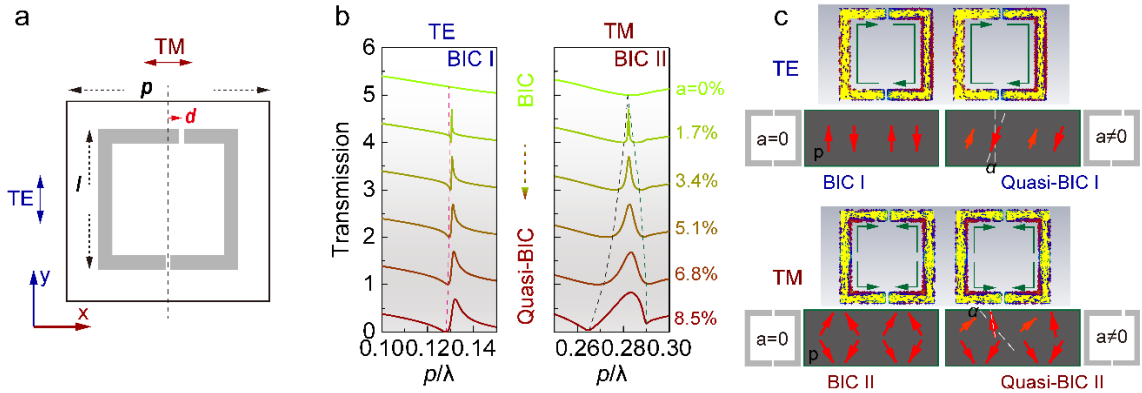

**Figure 3. Quasi-BICs induced by structural symmetry breaking. a.** Identical resonator geometry without $C_2$ symmetry by displacing the gap from the center with a distance '$d$'. A dimensionless parameter '$a$' is defined to numerically describe the asymmetry degree by $a = (l_1-l_2)/(l_1+l_2) \times 100\%$, where $l_1$ and $l_2$ indicate the total length of left and right branches of the resonator, respectively. **b.** Transmission spectra of the quasi-BICs at different asymmetry degrees at TE and TM polarizations. **c.** Analysis of BICs by electric dipole moments at TE and TM polarizations. Two identical resonators are shown in each scenario with surface current distributions and schematic of electric dipole arrangements. A slight perturbation on the dipoles would induce leakage of an ideal bound state.

In addition to tilting the incident wave vector, breaking the structural symmetry of the resonator itself is another pathway to access coupling of bound state to the normal incidence[22]. The bound state thus leaks (couples) to the incoming radiation giving rise to a quasi-BIC. As illustrated in Figure 3a, the geometric symmetry of a DSRR is broken by displacing one gap from the center by a distance 'd'. A dimensionless parameter is defined as $a = (l_1 - l_2)/(l_1 + l_2) \times 100\%$ to numerically measure the degree of asymmetry, where $l_1$ and $l_2$ indicate the total length of left and right branches of the resonator, respectively. The asymmetry degree is in fact a generalized parameter describing the leakage rate across the whole family of symmetry-protected BIC metamaterials[12, 23]. The transmission spectra are plotted in Figure 3b, where we capture the quasi-BIC modes with Fano- and EIT-like mode profiles at TE and TM polarizations, respectively. The linewidths of both quasi-BICs reveal strong dependence on the structural degree of asymmetry.

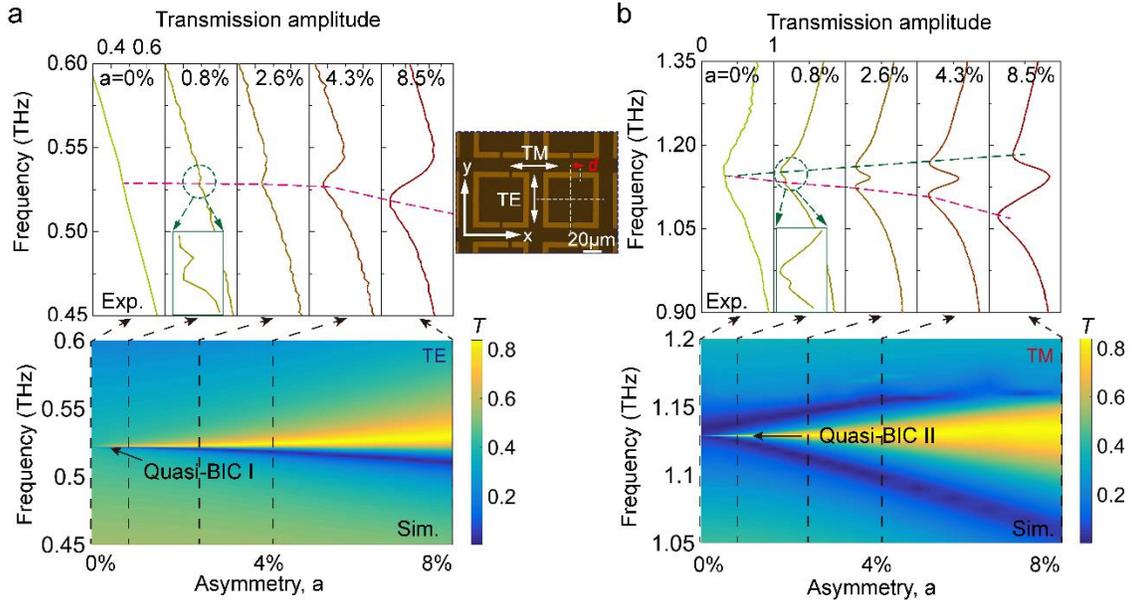

**Figure 4. Experimental observation of the quasi-BICs by tailoring the degree of structural asymmetry. a.** Bottom: simulated transmission amplitude spectra by varying 'a' from 0% to 8.5% with TE polarization. Disappearance of the Fano-like resonance occurs near $a = 0\%$ indicating a bound state without leakage. Top: experimental slices at five representative asymmetry degrees showing the spectral evolution, and inset graph shows the zoomed-in spectrum highlighting the ultra-narrow spectral feature at $a = 0.8\%$. The built-in graph shows a microscopic image of the sample with $a = 8.5\%$. **b.** Experimental and simulated transmission amplitude spectra with TM polarization, and inset graph shows the zoomed-in spectrum highlighting the ultra-narrow spectral feature at $a = 0.8\%$. The dashed lines in the top panels trace the evolution of transmission dips in the experimental spectra.

The transition from an ideal bound state to quasi-BIC is understood in terms of effective electric dipole moment ($p_{eff}$) as shown in Figure 3c. With a linearly polarized plane wave incidence, it can only couple to collective modes that have net electric components along *x* (*y*) axis with TM (TE) polarization due to the symmetry incompatibility. The electromagnetic multipoles can be extracted from the surface current distributions[24, 25]. At TE polarization, the antiphase currents are induced in the two wire branches of each DSRR which results in the electric dipoles with opposite polarities and equal magnitude at *a* = 0, and thus the net contribution is zero ($p_{net} = \sum_{n=1}^{N} p_{n,y}$, where *N* = 2 is the total number of current branches in each unit cell). Therefore, the resonant state is completely confined as a bound state in the cavity without leakage channel to incident plane wave. When the top gap is displaced (*a* ≠ 0), the dipole symmetry is broken in terms of the orientation as well as magnitude, which allows coupling of the mode to incident *y*-polarized radiation for quasi-BIC I. As a result, the BIC transitions to a quasi-BIC that reveals a symmetry-protected property. Similar observations occur for the BIC II. Instead of two branches of current flows, there are four mirror symmetric branches with TM polarization excitation leading to zero net electric dipole moment ($p_{net} = \sum_{n=1}^{N} p_{n,x}$, *N* = 4) at *a* = 0. Here, more branches of current distributions also explain the higher resonance frequency of the quasi-BIC II. Once the symmetry is perturbed by displacing the top gap, the orientation and magnitude of electric dipoles on the top branches are rearranged so that the bound state becomes leaky and couples to *x*-polarized incidence as a quasi-BIC.

The dual quasi-BICs are experimentally measured in the terahertz regime. We fabricated the DSRR samples by photolithography with geometrical parameters (in μm) *p* = 75, *l* = 60, *w* = 6, *g* = 3, and various asymmetry *d*, on an intrinsic silicon ($\varepsilon$ = 11.7, and $\rho$ > 5000 Ω m) substrate (inset graph of Figure 4). Aluminum (Al) is used to fabricate the resonators whose conductivity can be treated as DC with a large value ($\sigma = 3.6 \times 10^7$ S/m) at terahertz frequencies[26] so that it largely suppresses the Ohmic loss. The simulated (bottom panel) and experimental (top panel) transmission spectra are presented in Figure 4 by varying '*a*' from 0% to 8.5% (by displacing top gap with a distance *d*). The BICs are only spectrally visible by opening a leakage channel ($\gamma \neq 0$) from the ideal bound states by breaking the symmetry incompatibility at normal incidence. The transitions from BICs to quasi-BICs at TE and TM polarizations are observed in both simulations and experiments. Although a very slight perturbation of the $C_2$ symmetry results in a leakage of the BICs, it is beyond our measurement resolution limit if the spectral linewidth is narrower than 5 GHz (200 ps scan length). The smallest asymmetry degree where we could experimentally measure the transmission spectral features is at ~ 0.8%, and the extremely sharp quasi-BICs at TE and TM polarizations are highlighted in the zoomed-in graphs in Figure 4. The leakage rate is gradually enhanced at a larger '$a$' which reflects on the gradually broadened quasi-BIC spectral linewidth. We also observe the Fano- and EIT-like transmission profiles in the quasi-BIC scenarios that agree well with the simulated spectral features.

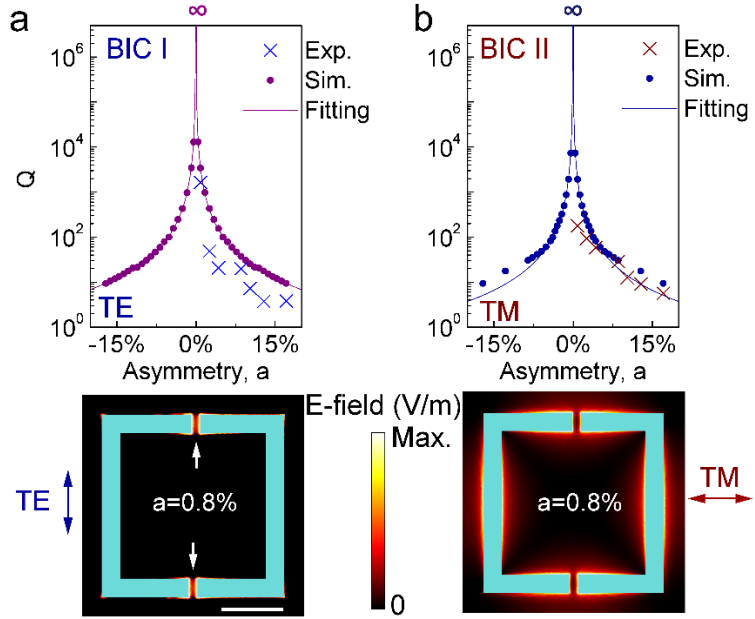

**Figure 5. Quality factors of the quasi-BICs.** The quality factors are extracted from simulations (dots, ideal scenarios of PEC) and experiments (crosses) by Fano fit for BIC I (**a**) and BIC II (**b**), respectively. Solid lines denote theoretical fit using an inverse square function. The bottom panel shows the local electric field distributions at TE and TM polarizations. Scale bar: 20 μm.

The most straightforward evidence of a BIC is the complete vanishing of leakage, and thereby giving rise to a divergent radiative quality factor that tends to infinity at a discrete point[12, 16]. In the subwavelength metamaterials, the considered frequencies of BICs are below the first-order diffraction limit, and only zeroth-order diffraction accounts for the leakage channel at quasi-BICs. Following the analysis of electric dipole distribution, the net electric dipole moment within each unit cell is written as $p_{net} = \sum_{n=1}^{N} p_{n,x(y)}$ for TM (TE) polarized incidence. Applying the perturbation theory at quasi-BICs, we obtain a simplified approximation of $\gamma_{rad} = \frac{\omega_0^2}{2cS}|p_{net}|^2$ where $c$ is the speed of light in vacuum, and $S$ is the area of a unit cell[12, 27]. As illustrated in Figure 3c, the net electric dipole moments depend on the rearrangement of dipole orientations and magnitude in a resonator, which is introduced by a slight gap displacement. Within the approximation of tiny perturbation, we could estimate $p_{net} \propto a$, and thus $\gamma_{rad} \propto a^2$ (refer to ref. 6 for rigorous numerical analysis). Therefore, the radiative $Q$ factor of a symmetry-protected quasi-BIC shows an inverse square dependence on '$a$' described by

$$Q_{rad} = m \cdot \frac{cS}{\omega_0 \cdot a^2} \tag{1}$$

where $m$ is a proportionality constant. We note that this inverse square relation remains valid for the whole family of symmetry-protected quasi-BICs as long as the quasi-BIC frequency is within the zero-order diffraction limit.

In order to map the diverging trajectory of $Q$ factors, we retrieve $Q_{tot}$ by using Fano line shape fit for simulated and measured transmission spectra, and plot them as dots and crosses in Figure 5, respectively. In simulations, Ohmic loss is ignored, and $Q_{tot} = Q_{rad}$. The extracted $Q$ factors of quasi-BICs show a clear diverging trend to infinity near $a = 0\%$. The diverging trajectories are also validated by the theoretical model by using the inverse square equation to fit the data. We obtain an excellent fit as shown in the graphs. The slight disagreement at larger asymmetry is due to the deviation from the tiny perturbation assumption in Equation 1. Total $Q$ factors of experimental spectra with '$a$' varying from 0% to 17.1% are also extracted and plotted as crosses in Figure 5. The values of $Q$ factors from experiments are overall lower than those of simulations and theoretical model since both radiative and non-radiative (Ohmic) losses are considered in experiments to estimate the total $Q$. In addition, inevitable scattering losses due to the rough surfaces and finite area of samples also account for the decrease in the total $Q$ factors. Nevertheless, the experimental $Q$ factors show good agreement with the diverging trajectories following the quasi-BIC features. We note that the largest $Q$ values that could be experimentally measured are limited by instrument resolution and signal-to-noise ratio. Therefore, we have experimentally demonstrated the dual BIC features of a subwavelength metamaterial, and observed the optical quasi bound states in a symmetry-compatible radiation continuum.

The significant advantage of the metamaterial-based BICs is the tight confinement of localized field in the subwavelength scale. As an example, the electric field distributions of quasi-BICs are shown in the bottom panel of Figure 5 with $a = 0.8\%$. At quasi-BIC I, most incident terahertz radiation at $\lambda = 574$ μm is tightly confined in a small area of split gaps in a resonator (3 μm in width), which could be further reduced to nanoscale with a deep subwavelength mode volume[28]. At quasi-BIC II, although the field is confined in the vicinity of resonator arms with a relative larger volume, it is still at the subwavelength scale. The BIC-inspired extremely high quality factors and metamaterial-inspired deep subwavelength mode volume would significantly enhance the Purcell factor ($\propto Q/V$) for high-performance lasing and sensing applications.

In summary, we have experimentally demonstrated the *dual bound states* in the continuum in a subwavelength metamaterial with $C_2$ symmetry. Eigenmode analysis reveals the existence of the polarization-dependent, symmetry-protected BICs. Terahertz experiments verified the dual quasi-BICs by opening a leakage channel to probe the far-field spectral features. We also observe the diverging radiative quality factors and unveil the inverse square dependence on structural asymmetry. The subwavelength metamaterial provides an ideal platform to access BIC for strong light-matter interactions. The BIC resonant systems could also be exploited as a general strategy in different applications such as narrow-band filters and energy-efficient modulators for terahertz wireless communications.

**Supporting Information**

Supporting Information is available from the Wiley Online Library or from the author.

**Acknowledgements**

The authors acknowledge the research funding support from the Singapore Ministry of Education (MOE2017-T2-1-110, and MOE2016-T3-1-006(S)) and the National Research Foundation (NRF), Singapore and Agence Nationale de la Recherche (ANR), France (grant No. NRF2016-NRF-ANR004).

**Conflict of Interest**

The authors declare no conflict of interest.